\documentclass[aps,onecolumn,9pt,tightenlines,notitlepage,nofootinbib]{revtex4-1}
\usepackage[utf8]{inputenc}

\usepackage{amsmath}
\usepackage{amsfonts}
\usepackage{amssymb}
\usepackage{latexsym}
\usepackage{amstext}

\usepackage{graphicx}
\usepackage{multirow}

\newcommand{\be}{\begin{equation}}
\newcommand{\ee}{\end{equation}}
\newcommand{\bea}{\begin{eqnarray} \nonumber }
\newcommand{\eea}{\end{eqnarray}}
\newcommand{\bi}{\begin{itemize}}
\newcommand{\ei}{\end{itemize}}

\begin{document}
\title{Election turnout statistics in many countries: similarities, differences, and a diffusive field model for decision-making}
\author{Christian Borghesi$^1$, Jean-Claude Raynal$^2$ and Jean-Philippe Bouchaud$^3$}
\email{borghesi@msh-paris.fr ; raynal@ehess.fr ; jean-philippe.bouchaud@cea.fr}
\affiliation{$1$: Centre d'Analyse et de Math\'ematique Sociales (CAMS-EHESS), 190-198, avenue de France, 75013 Paris, France\\
$2$: EHESS, Division Histoire, 190-198, avenue de France, 75013 Paris, France\\
$3$: Capital Fund Management, 6-8 Bd Haussmann, 75009 Paris, France}
\date{\today}
\begin{abstract}
We study in details the turnout rate statistics for 77 elections in 11 different countries. We show that the empirical results established in a previous paper
for French elections appear to hold much more generally. We find in particular that the spatial correlation of turnout rates decay logarithmically with distance 
in all cases. This result is quantitatively reproduced by a decision model that assumes that each voter makes his mind as a result of three influence terms: one totally idiosyncratic component, 
one city-specific term with short-ranged fluctuations in space, and one long-ranged correlated field which propagates diffusively in space. A detailed analysis reveals several interesting features: 
for example, different countries have different degrees of local heterogeneities and seem to be characterized by a different propensity for individuals to conform to the cultural norm. We furthermore 
find clear signs of herding (i.e. strongly correlated decisions at the individual level) in some countries, but not in others.
\end{abstract}
\maketitle

\section{Introduction}

Empirical studies and models of election statistics have attracted considerable attention in the recent physics literature, see e.g. \cite{costa_filho_scaling_vot,lyra_bresil_el,gonzalez_bresil_inde_el,fortunato_universality,daisy-model,growth_model_vote,araripe_role_parties,araujo_tactical_voting,universality_candidates}.
In \cite{diffusive-field}, the present authors have studied the statistical regularities of the electoral turnout rates, based on spatially resolved data from 13 French elections since 1992. 
Two striking features emerged from our analysis: first, the distribution of the logarithmic turnout rate $\tau$ (defined precisely below) was found to be remarkably stable over all 
elections, up to an election dependent shift. Second, the spatial correlations of $\tau$ was found to be well approximated by an affine function of the {\it logarithm} of the distance 
between two cities. Based on these empirical results, we proposed that the behaviour of individual agents is affected by a space dependent ``cultural field'', that encodes a local bias
in the decision making process (to vote or not to vote), common to all inhabitants of a given city. The cultural field itself can be decomposed into an idiosyncratic part, with short range
correlations, and a slow, long-range part that results from the diffusion of opinions and habits from one city to its close-by neighbours. We showed in particular that this local propagation 
of cultural biases generates, at equilibrium, the logarithmic decay of spatial correlations that is observed empirically \cite{diffusive-field}. 

The aim of the present note is to provide additional support to these rather strong statements, using a much larger set of elections from different countries in the world. We discuss in more 
depth the approximate universality of the distribution of turnout rates, and show that some systematic effects in fact exist, related in particular, to the size of the cities. We also confirm that the logarithmic decay of the spatial correlations approximately holds for all countries and all elections, with parameters compatible with our diffusive field model. The relative importance 
of the idiosyncratic, city dependent contribution and of the slow diffusive part is however found to be strongly dependent on countries. We also confirm the universality of the logarithmic 
turnout rate for different elections, for different regions or for different cities, provided the mean and the width of the distribution is allowed to depend on the city size. 
Overall, our empirical analysis provides further support to the binary logit model of decision making, with a space dependent mean (the cultural field mentioned above).

\section{Data and Observables}

We have analyzed the turnout rate at the scale of municipalities for 77 elections, from 11 different countries. For some countries, the number of different elections is substantial:  
22 from France (Fr, $\approx36000$ municipalities in mainland France) \cite{data-mun-fr}, 13 from Austria (At, $\approx2400$ municipalities) \cite{data-mun-at}, 
11 from Poland (Pl, $\approx2500$ municipalities) \cite{data-mun-pl},  7 from Germany (Ge, $\approx12000$ municipalities) \cite{data-mun-ge}, while for others 
we have less samples: 5 from Canada (Ca, $\approx7700$ municipalities) \cite{data-mun-ca}, 4 from Romania (Ro, $\approx3200$ municipalities) \cite{data-mun-ro}, 4 from Spain 
(Sp, $\approx8000$ municipalities in mainland Spain) \cite{data-mun-sp}, 4 from Italy (It, $\approx7200$ municipalities in mainland Italy) \cite{data-mun-it}, 
3 from Mexico (Mx, $\approx2400$ municipalities) \cite{data-mun-mx}, 3 from Switzerland (CH, $\approx2700$ municipalities) \cite{data-mun-ch} and 1 from Czech Republic (Cz, $\approx6200$ municipalities)\cite{data-mun-cz}. More details on the nature of these elections and some specific issues are given in Appendix.

For each municipality and each election, the data files give the total number of registered voters $N$ and the number of actual voters $N_+$, 
from which one obtains the usual turnout rate $\pi=N_+/N \in [0,1]$. For reasons that will become clear, we will instead consider in the following the logarithmic turnout rate (LTR) $\tau$, 
defined as:
\be 
\label{etau} 
\tau:=\ln(\frac{\pi}{1-\pi})~, 
\qquad 
\tau \in ]-\infty,+\infty[.
\ee
Because we know the geographical location of each city, the knowledge of $\tau$ for each city enables us to create a map of the field $\tau(\vec r)$ and study its spatial correlations. 

\section{Statistics of the local turnout rate}

Whereas the average turnout rate is quite strongly dependent on the election (both on time and on the type of election -- local, presidential, referendum, etc.), 
the distribution of the shifted LTR $\tau - \langle \tau \rangle$ was found to be remarkably similar for the 13 French elections studied in \cite{diffusive-field}.\footnote{The notation $\langle \dots \rangle$ means a flat average over all cities (i.e. not
weighted by the population $N$ of the city).} The LTR standard-deviation, skewness and kurtosis were found to be very similar between different elections. 
The distribution $P(u)$ of the shifted and rescaled LTR,
\be
u = \frac{\tau - \langle \tau \rangle}{\sigma},\quad {\rm with} \quad \sigma^2 = \langle \tau^2 \rangle - \langle \tau \rangle^2
\ee
was found to be very close in the Kolmogorov-Smirnov (KS) sense.

\begin{figure}
\begin{minipage}[c]{0.6\linewidth}
\includegraphics[width=9cm, height=6cm, clip=true]{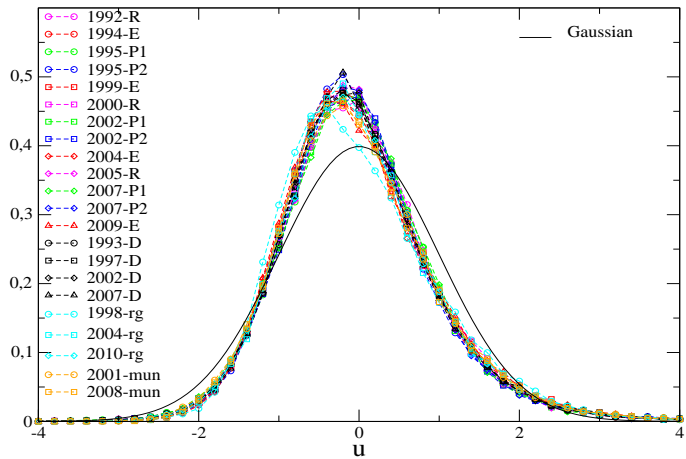}
\end{minipage}\hfill
\begin{minipage}[c]{0.35\linewidth}
\caption{\small Probability distribution of the rescaled variable $u$ over all {\it communes} for France. A standardized Gaussian is also shown. 
We use the same symbols and color codes for the French elections throughout this paper.}
\label{fPu}
\end{minipage}
\end{figure}

We have extended this analysis to the 9 new election data in France, and to all new countries mentioned above. For France, the {\it Elections Municipales}
(election of the city mayor), not considered in \cite{diffusive-field}, have a distinctly larger standard deviation than national elections. However, $P(u)$
is again found to be similar for all the French elections, except the {\it R\'egionales} of 1998 and 2004. These happen to be coupled with other local elections in half municipalities, which 
clearly introduces a bias. The distributions $P(u)$ for all elections in France are shown in Fig.~\ref{fPu} and compared to a Gaussian variable. The distribution is clearly non Gaussian, with a positive skewness equal to $\approx 1.1$ and a kurtosis equal $\approx 4.8$. A more precise analysis consists in computing the KS distances between each pair of elections. We recall here that a KS distance 
of $d_{KS}=1$ corresponds to a $\approx 20 \%$ probability that the two tested distribution coincide, while $d_{KS}=1.6$ corresponds to a $\approx 1 \%$ probability. Removing the {\it R\'egionales}, we find that the KS distance $d_{KS}$ averaged over all pairs of elections is equal to $1.49$, with a standard deviation of $0.47$. These numbers are slightly too large to ascertain that the distributions are 
exactly the same since in that case the average $d_{KS}$ should be equal to $\sqrt{\pi/2} \times \ln 2 \approx 0.87$. On the other hand, these distances are not large either (as visually clear from Fig.~\ref{fPu}), meaning that while systematic differences between elections do exist, they are quite small. We will explain below a possible origin for these differences.
 
The same analysis can be done for all countries separately; as for France, we find that $P(u)$ for different elections are all similar, except for Germany for which $\langle d_{KS} \rangle=3$ 
-- see Table~\ref{tks+stat}, where we show the mean and the standard-deviation of KS distances between elections of a given country, and of the skewness and kurtosis of the distributions $P(u)$ in a given country. Note that the
values of $\langle d_{KS} \rangle$ are close to $0.87$ for Italy and Poland. On the other hand, these distributions
is clearly found {\it not} to be identical across different countries. Table~\ref{tks-superdistrib} shows the matrix of KS distances between countries ``super-distributions".\footnote{A ``super-distribution" of $\tau$ of a country is obtained by aggregating the appropriately shifted LTR distributions over all  ``compatible'' elections. Compatible elections have roughly the same distribution $P(\tau-\langle\tau\rangle)$, i.e. {\it without} normalization by its standard-deviation. They are chosen as follows: for Canada and Poland all elections; for France all pure national elections (nor combined with local elections, i.e. all elections apart from 1998-rg, 2004-rg, 2001-mun and 2008-mun); for Mexico 2003-D and 2009-D; for Germany 2005-D and 2009-D; all Chamber of Deputies (D) elections for Austria, Spain, Italy and Switzerland; and for Romania, all elections apart from its European Parliament election (see Appendix for more details).} The values of $d_{KS}$ are all large, except for the pairs Fr-Cz, Fr-CH, Sp-CH, Sp-Ro and 
CH-Cz.

\begin{table}
\begin{tabular}{|l c l l | l c l l|}
\hline
Country & $d_{KS}$ & skewness & kurtosis & Country & $d_{KS}$ & skewness & kurtosis \\
\hline
\multirow{2}*{Fr} & 1.49$\pm$0.47 &	1.08$\pm$0.15 & 4.8$\pm$1.3	& \multirow{2}*{At} & 1.44$\pm$0.54 	& 0.10$\pm$0.38 & 0.53$\pm$0.81 \\
			& (1.42$\pm$0.45) &	(1.10$\pm$0.14) & (5.1$\pm$0.9)	& 					& (0.93$\pm$0.19) 	& (-0.13$\pm$0.21) & (0.54$\pm$0.43) \\
\hline
\multirow{2}*{Pl} & 0.80$\pm$0.20 &	0.12$\pm$0.26 & 0.38$\pm$0.42 & \multirow{2}*{Ge} & 3.0$\pm$1.1 	& 0.48$\pm$0.30 & 1.6$\pm$0.9 \\
 				& (0.80$\pm$0.20) &	(0.12$\pm$0.26) & (0.38$\pm$0.42) &					& (0.81)		& (0.20$\pm$0.05) & (1.53$\pm$0.04) \\
\hline
\multirow{2}*{Sp} & 1.78$\pm$0.68 &	0.27$\pm$0.25 & 1.8$\pm$1.1 & \multirow{2}*{It} & 0.70$\pm$0.09 	& -0.45$\pm$0.11 & 1.01$\pm$0.02 \\
				& (1.24) 		  & (0.07$\pm$0.21)	& (2.5$\pm$1.2) &				& (0.68)			& (-0.45$\pm$0.15) & (1.01$\pm$0.003) \\
\hline
\multirow{2}*{CH} & 1.67$\pm$0.43 &	0.51$\pm$0.08 & 1.4$\pm$1.4	& \multirow{2}*{Mx} & 1.28$\pm$0.35		& 0.32$\pm$0.09 & 1.1$\pm$0.8 \\
				&			 	  &	(0.47)		& (2.9)			&					& (1.19)			& (0.35$\pm$0.11) & (1.6$\pm$0.3) \\
\hline
\multirow{2}*{Ca} & 1.23$\pm$0.39 &	-0.40$\pm$0.39 & 4.4$\pm$0.9 & \multirow{2}*{Ro} & 1.06$\pm$0.39	& 0.05$\pm$0.43 & 1.5$\pm$0.4 \\
				& (1.23$\pm$0.39) &	(-0.40$\pm$0.39) & (4.4$\pm$0.9) &				& (0.95$\pm$0.36)	& (-0.14$\pm$0.25) & (1.6$\pm$0.4) \\
\hline
\end{tabular}
\caption{\small Mean and standard-deviation of KS distances ($d_{KS}$) between all pairs of elections within each country. Mean and standard-deviation of skewness and kurtosis of distributions of $\tau$ over all municipalities is also given for each country. In parentheses, the same measures but restricted to compatibles elections in each country.}
\label{tks+stat}
\end{table}

\begin{table}
\begin{tabular}{|l|c c c c c c c c c c c|}
\hline
 & Fr & At & Pl & Ge & Sp & It & Cz & Mx & CH & Ca & Ro \\
\hline
Fr & 0 & 5.01 & 5.61 & 8.00 & 2.28 & 6.13 & 0.93 & 2.18 & 0.83 & 6.72 & 2.66 \\
At & 5.01 & 0 & 1.62 & 1.49 & 2.43 & 1.58 & 3.24 & 2.31 & 2.25 & 4.60 & 1.57 \\
Pl & 5.61 & 1.62 & 0 & 2.32 & 2.41 & 3.12 & 3.16 & 1.83 & 2.06 & 6.62 & 1.99 \\
Ge & 8.00 & 1.49 & 2.32 & 0 & 3.74 & 1.73 & 4.84 & 2.81 & 2.83 & 7.15 & 2.85 \\
Sp & 2.28 & 2.43 & 2.41 & 3.74 & 0 & 3.17 & 1.71 & 2.19 & 1.11 & 3.53 & 0.95 \\
It & 6.13 & 1.58 & 3.12 & 1.73 & 3.17 & 0 & 3.65 & 3.13 & 2.58 & 4.62 & 2.05 \\
Cz & 0.93 & 3.24 & 3.16 & 4.84 & 1.71 & 3.65 & 0 & 2.12 & 0.58 & 2.45 & 1.94 \\
Mx & 2.18 & 2.31 & 1.83 & 2.81 & 2.19 & 3.13 & 2.12 & 0 & 1.87 & 4.06 & 1.95 \\
CH & 0.83 & 2.25 & 2.06 & 2.83 & 1.11 & 2.58 & 0.58 & 1.87 & 0 & 1.44 & 1.39 \\
Ca & 6.72 & 4.60 & 6.62 & 7.15 & 3.53 & 4.62 & 2.45 & 4.06 & 1.44 & 0 & 2.78 \\
Ro & 2.66 & 1.57 & 1.99 & 2.85 & 0.95 & 2.05 & 1.94 & 1.95 & 1.39 & 2.78 & 0 \\
\hline
\end{tabular}
\caption{\small Kolmogorov-Smirnov distance between different ``super-distributions''.}
\label{tks-superdistrib}
\end{table}

\begin{figure}
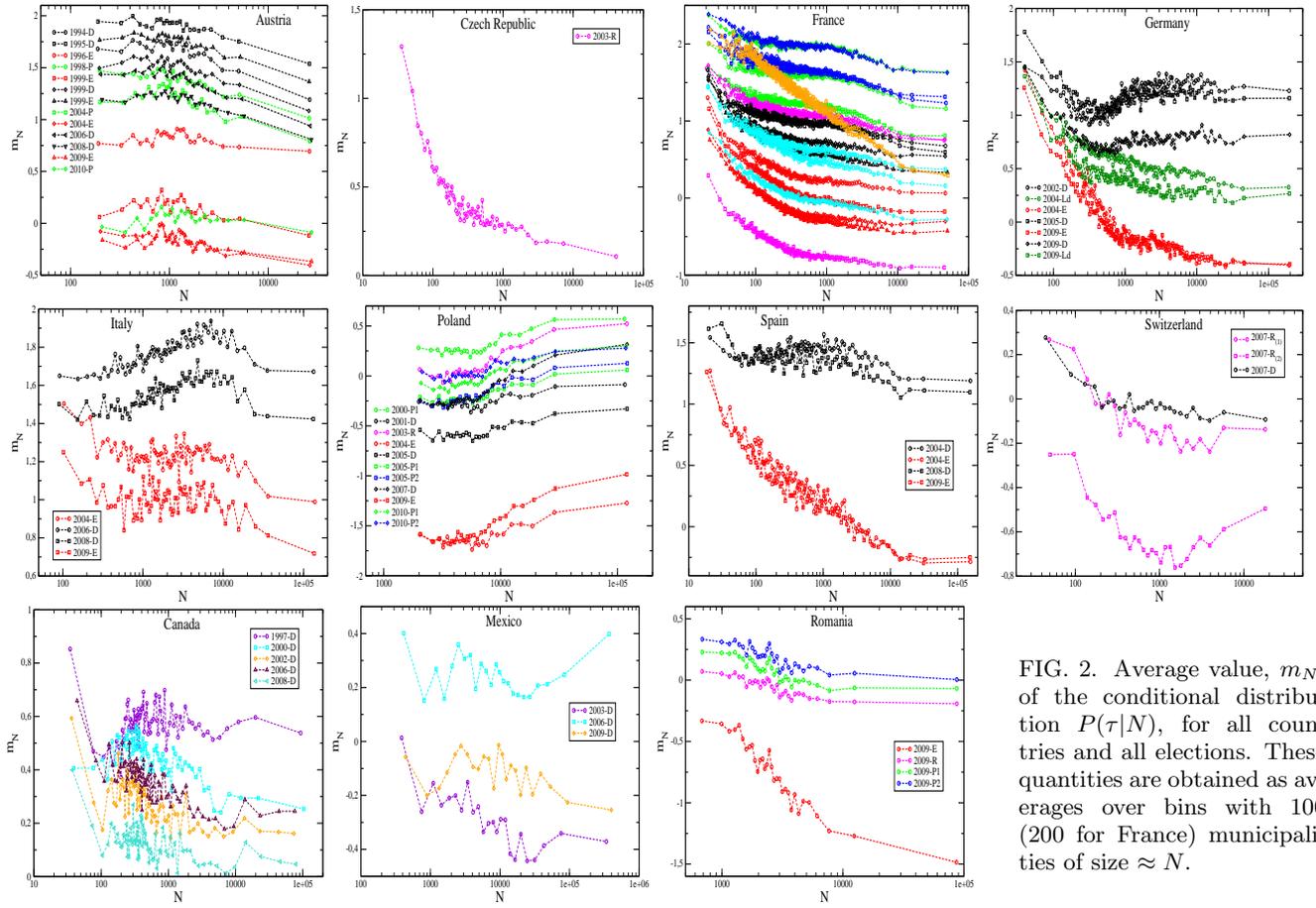

\includegraphics[width=4.25cm, height=4cm, clip=true]{tau-N-at.eps}
\includegraphics[width=4.25cm, height=4cm, clip=true]{tau-N-cz.eps}
\includegraphics[width=4.25cm, height=4cm, clip=true]{tau-N-fr.eps}
\includegraphics[width=4.25cm, height=4cm, clip=true]{tau-N-ge.eps}\\
\includegraphics[width=4.25cm, height=4cm, clip=true]{tau-N-it.eps}
\includegraphics[width=4.25cm, height=4cm, clip=true]{tau-N-pl.eps}
\includegraphics[width=4.25cm, height=4cm, clip=true]{tau-N-sp.eps}
\includegraphics[width=4.25cm, height=4cm, clip=true]{tau-N-ch.eps}
\begin{minipage}[c]{0.75\linewidth}
\includegraphics[width=4.25cm, height=4cm, clip=true]{tau-N-ca.eps}
\includegraphics[width=4.25cm, height=4cm, clip=true]{tau-N-mx.eps}
\includegraphics[width=4.25cm, height=4cm, clip=true]{tau-N-ro.eps}
\end{minipage}\hfill
\begin{minipage}[c]{0.23\linewidth}
\caption{\small Average value, $m_N$, of the conditional distribution $P(\tau|N)$, for all countries and all elections.
These quantities are obtained as averages over bins with 100 (200 for France) municipalities of size $\approx N$.}
\label{fmN}
\end{minipage}
\end{figure}

In order to understand better these results, one should first realize that the statistics of the LTR does in fact strongly depend on the size of the cities.
This was already pointed out in \cite{diffusive-field,these}. For example, the average LTR for all cities of size $N$ (within a certain interval), that we
denote as $\langle \tau \rangle_N \equiv m_N$, is distinctly $N$ dependent, see Fig.~\ref{fmN}. In most cases, the average turnout rate is large in small 
cities and declines in larger cities, with notable exceptions: for example, the trend is completely reversed in Poland, with more complicated patterns for 
parliament elections in Italy or Germany. Similarly, the standard-deviation of $\tau$, $\sigma_N$, also depends quite strongly on $N$ (see Figs \ref{fsigma2N-fr} and \ref{fsigma2N} below).

However, the distribution $Q_N(v)$ of the rescaled variable $v =(\tau - m_N)/\sigma_N$ over all cities of size $N$ for each election can be considered to be {\it universal} from a KS point of view, 
both within the same country for different $N$ but now also, when $N$ is large enough, across different countries. For example, the average KS distance between distributions corresponding to different ranges of $N$ in France is equal to $0.58$, with standard-deviation $0.12$. These numbers are respectively $0.72 \pm 0.20$, $0.58 \pm 0.13$ and $0.87 \pm 0.36$ for Italy, Spain and Germany.\footnote{We have excluded the smallest cities, $N  < 200$, 
that are have a distinctly larger KS distance with other cities -- see below. Bins, ranked according to the municipality size $N$ contain each around 500 municipalities.} In Table~\ref{tks-superdistrib-N}, we show for different bins of $N$ values the mean and standard-deviation KS distance between countries, 
illustrating that all distributions are statistically compatible, at least when $N$ is large enough. 

Now, even if $Q_N(v)$ is universal and equal to $Q^*(v)$, $P(u)$ will reflect the country-specific (and possibly election-specific) shapes of $m_N$ and $\sigma_N$, and the country-specific distribution of city sizes, $\rho(N)$. Indeed, one has:
\be
P(\tau) = \sum_{N} \rho(N) \, Q^*\left(\frac{\tau - m_N}{\sigma_N}\right),
\ee
which has no reason whatsoever to be universal. But since for a given country the dependence on $N$ of $m_N,\sigma_N$ and $\rho(N)$ tends to change only weakly in time, the 
approximate universality of $P(u)$ for a given country follows from that of $Q_N(v)$. In fact, French national elections can be grouped into two
families, such that the dependence of $m_N$ on $N$ is the same within each family but markedly different for the two families (see next section and Fig.~\ref{fmN-fr} below). 
Restricting the KS tests to pairs within each families now leads to an average KS distance of $\approx 1.25$ with a standard deviation $\approx 0.4$ (identical for the two families), 
substantially smaller than $d_{KS}= $ from Table~\ref{tks+stat}. This goes to show that the election specific shape of $m_N$ is indeed partly responsible for the weak non-universality of $P(u)$.

\begin{table}
\begin{tabular}{| l | c c c c c|}
\cline{2-6}
\multicolumn{1}{l|}{} & ~~$1000\leq N < 2000$~~ & ~~$2000\leq N < 4000$~~ & ~~$4000\leq N < 8000$~~ & ~~$8000\leq N < 16000$~~  & ~~$16000\leq N$\\
\hline
$d_{KS}$ & 1.47$\pm$0.77 & 1.38$\pm$0.65 & 0.94$\pm$0.48 & 0.91$\pm$0.46 & 0.95$\pm$0.48\\
\hline
\end{tabular}
\caption{\small Mean and standard-deviation over all pairs of countries of the KS distance $d_{KS}$ between the aggregated $Q_N(v)$ distributions in each country, for different values of $N$.}
\label{tks-superdistrib-N}
\end{table}

Zooming in now on details, we give in Table~\ref{tks-gauss} the KS distance between $Q_N(v)$ aggregated over all elections of a country and a normalized Gaussian, for different ranges of $N$ and different countries. 
The skewness and kurtosis of the distribution $Q^*(v)$ and the KS distance to a Gaussian, aggregated over all $N$, are given in Table~\ref{tQ-aggregated}-a for different countries, 
and aggregated over countries for fixed $N$ in Table~\ref{tQ-aggregated}-b. Two features emerge from these Tables: 
\begin{itemize}
\item While for some countries (Cz, Sp, Mx) the deviation of $Q_N(v)$ from a Gaussian appear small (both measured by KS or by the skewness and kurtosis), such an assumption is clearly unacceptable for Italy and Germany, 
for which the KS distance is large for all $N$ (see Table~\ref{tks-gauss}) and a substantial negative skewness can be measured. 
Furthermore, the aggregated distribution (over all $N$) is clearly incompatible with a Gaussian except in the 
Czech Republic, Spain and Mexico.
\item There is an interesting systematic $N$ dependence of the distance to a Gaussian, which is on average smaller for larger $N$s, and maximum for small cities. 
This suggests that although the KS tests is unable to distinguish strongly the $Q_N(v)$ for different $N$, there is in fact a systematic evolution for
which we provide an argument below. In fact, as clearly seen in Table~\ref{tks-superdistrib-N}, the average KS distance between the $Q_N$ of different countries is also systematically smaller as $N$ increases.  
\end{itemize}

\begin{table}
\begin{tabular}{| l c c c c c |}
\hline
Country & ~~$1000\leq N < 2000$~~ & ~~$2000\leq N < 4000$~~ & ~~$4000\leq N < 8000$~~ & ~~$8000\leq N < 16000$~~  & ~~$16000\leq N$\\
\hline
Fr & 2.50 & 2.15 & 1.18 & 0.71 & 0.86\\
At & 2.03 & 1.82 & 0.76 & 0.98 & 1.58\\
Pl & 0.45 & 1.45 & 0.89 & 1.40 & 1.20\\
Ge & 1.75 & 2.78 & 2.55 & 2.49 & 3.08\\
Sp & 0.70 & 0.83 & 0.71 & 0.63 & 0.69\\
It & 2.69 & 3.74 & 3.11 & 2.32 & 0.88\\
Cz & 0.63 & 0.73 & 0.55 & 0.37 & 0.61\\
Mx & 1.50 & 0.79 & 0.55 & 0.97 & 0.48\\
CH & 1.38 & 1.49 & 0.65 & 0.69 & 0.44\\
Ca & 3.48 & 1.09 & 0.60 & 0.53 & 0.59\\
Ro & 1.73 & 1.48 & 1.14 & 0.63 & 0.92\\
\hline
\end{tabular}
\caption{\small KS distance between $Q_N(v)$ and a normalized Gaussian for different ranges of $N$ and for different countries.}
\label{tks-gauss}
\end{table}

\begin{table}
\begin{minipage}[c]{0.475\linewidth}
\begin{tabular}{|l c c c |}
\hline
Country & $d_{KS}$ & ~~skew~~ & ~~kurt~~\\
\hline
Fr & 2.55 & -0.02 & 0.31 \\
At & 2.63 & -0.05 & 0.15 \\
Pl & 2.13 & 0.18 & 0.58 \\
Ge & 4.09 & -0.21 & 0.05 \\
Sp & 1.03 & -0.16 & 0.41 \\
It & 5.61 & -0.67 & 0.79 \\
Cz & 0.83 & -0.32 & 0.30 \\
Mx & 1.21 & 0.12 & -0.06 \\
CH & 1.85 & 0.24 & 0.88 \\
Ca & 2.93 & -0.75 & 2.14 \\
Ro & 2.36 & -0.06 & 1.25 \\
\hline
\multicolumn{4}{c} {Tab.~\ref{tQ-aggregated}-a}\\
\end{tabular}
\end{minipage}\hfill
\begin{minipage}[c]{0.475\linewidth}
\begin{tabular}{|l c c c |}
\hline
Range of $N$ & $d_{KS}$ & ~~skew~~ & ~~kurt~~\\
\hline
$1000\leq N<2000$ & 2.25 & -0.07 & 0.43 \\
$2000\leq N<4000$ & 3.50 & -0.12 & 0.44 \\
$4000\leq N<8000$ & 2.90 & -0.12 & 0.42 \\
$8000\leq N<16000$ & 1.74 & -0.13 & 0.31 \\
$16000\leq N$ & 1.74 & -0.19 & 0.43 \\
\hline
\multicolumn{4}{c} {Tab.~\ref{tQ-aggregated}-b}\\
\end{tabular}
\end{minipage}
\caption{\small KS distance ($d_{KS}$) to a standardized Gaussian, and low-moment skewness (skew) and kurtosis (kurt) of aggregated distributions $Q^*(v)$. Tab.~\ref{tQ-aggregated}-a: data are aggregated over all $N$ for each country. Tab.~\ref{tQ-aggregated}-b: data are aggregated over all countries for fixed $N$.}
\label{tQ-aggregated}
\end{table}

\section{A theoretical canevas}

In order to delve deeper into the meaning of the above results, we need a theoretical framework. In \cite{diffusive-field}, we proposed to extend the
classical theory of choice to account for spatial heterogeneities. A registered voter $i$ makes the decision to vote ($S_i=1$) or not ($S_i=0$) on a given election. 
We can view this binary decision as resulting from a continuous and unbounded variable $\varphi_i \in ]-\infty,+\infty[$ that we called {\it intention} 
(or propensity to vote). The final decision depends on the comparison between $\varphi_i$ and a {\it threshold value} $-\Phi_{th}$: 
$S_i=1$ when $\varphi_i > -\Phi_{th}$, and $S_i=0$ otherwise. 
In~\cite{diffusive-field}, the intention $\varphi_i(t)$ of an agent at time $t$ who lives in a city $a$, located in the vicinity of $\vec R_a$, was decomposed as:
\be 
\label{evarphii} \varphi_i(t) = \epsilon_i(t) + \phi(\vec R_a,t) + \mu_a(t);
\ee
where $\epsilon_i(t)$ is the instantaneous and idiosyncratic contribution to the intention that is specific to voter $i$, and 
$\phi(\vec R,t)$ and $\mu_a(t)$ are fields that locally bias the decision of agents living in the same area. The first field $\phi$ is assumed to
be smooth (i.e. slowly varying in time and space), as the result of the local influences of the surroundings. This is what we called a ``cultural 
field'', that transports (in space) and keeps the memory (in time) of the collective intentions. The second field $\mu_a$, on the other hand, is
city- and election-specific, and by assumption has small inter-city correlations. It reflects all the elements in the intention that depend on 
the city: its size, the personality of its mayor, the specific importance of the election that might depend on the 
socio-economic background of its inhabitants, as well as the fraction of them who recently settled in the city, etc. 
(See \cite{diffusive-field} for a more thorough discussion of Eq.~(\ref{evarphii}).)

Consider now $N$ agents living in the same city, i.e. with under the influence of same field values $\phi$ and $\mu$. The turnout rate $\pi$ is by definition: 
\be
\pi=\frac{1}{N} \sum_{i=1}^{N} S_i.
\ee
For $N$ sufficiently large, and if the agents make {\it independent decisions}, the Central Limit Theorem tells us that:
\be
\pi \approx  p + \sqrt{\frac{p(1-p)}{N}} \xi,
\ee
where $p={\cal P}(\varphi > -\Phi_{th})$ is the probability that the conviction of the voter is strong enough, and $\xi$ is a standardized Gaussian noise.
If, on the other hand, agents make correlated decisions (for example, everybody in a family decides to vote or not to vote under the influence of a strong leader), one expects 
the variance of the noise term to increase by a certain ``herding'' factor $h \geq 1$, which measures the average size of strongly correlated groups. Therefore we
will write more generally:
\be
\pi \approx  p + \sqrt{\frac{h\,p(1-p)}{N}} \, \xi.
\ee

Following a standard assumption in Choice Theory, we take the idiosyncratic $\epsilon$'s to have a logistic distribution with zero mean and 
standard-deviation $\Sigma$, in which case the expression of $p$ becomes:
\be 
\label{ep} 
p=\frac{1}{1+\exp(-\frac{\phi+\mu+\Phi_{th}}{\Sigma})}.
\ee
This allows one to obtain a very simple expression for the LTR $\tau$:
\be
\label{etaua} \tau = \ln(\frac{\pi}{1-\pi}) \approx \beta \cdot \left(\phi + \mu + \Phi_{th} \right) + \sqrt{\frac{h}{Np(1-p)}} \, \xi,
\ee
where $\beta \equiv 1/\Sigma$. Therefore, in this model, the statistics of $\tau$ directly reflects that of the cultural and idiosyncratic fields.

\begin{figure}
\begin{minipage}[c]{0.6\linewidth}
\includegraphics[width=9cm, height=6cm, clip=true]{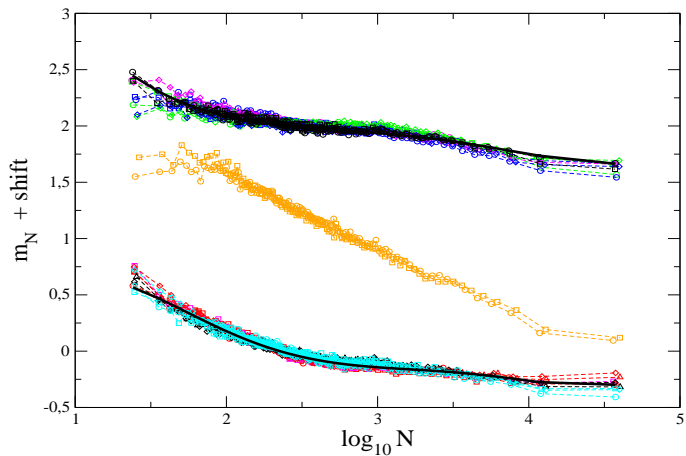}
\end{minipage}\hfill
\begin{minipage}[c]{0.35\linewidth}
\caption{\small Shifted $\langle \tau \rangle_N = m_N$ as a function of $N$ for French elections. Three families of elections clearly appear. a) Top curves: ``important'' national elections (Presidential, Referendums, Parliament); b) Bottom curves: less important national elections (European, {\it R\'egionales}); and c) Middle curves: {\it Municipales} (see text). Each point comes from the average over around 200 {\it communes} of size $\approx N$.}
\label{fmN-fr}
\end{minipage}
\end{figure}

Let us work out some consequences of the above decomposition, and how they relate to the above empirical findings. 
Since the cultural field $\phi$ is by definition not attached to a particular city, it is reasonable to assume that $\phi$ and $\beta$ are uncorrelated. 
Without loss of generality, one can furthermore set $\langle \phi \rangle = \langle \mu \rangle = 0$. Therefore:
\be
\langle \tau \rangle_N = m_N = \langle \beta \rangle_N \Phi_{th} + \langle \beta \mu \rangle_N.
\ee
Two extreme scenarios can explain the $N$ dependence of $m_N$: one is that the dispersion term $\langle \beta \rangle$ is strongly $N$ dependent 
while the statistics of $\mu$ is $N$ independent, the other is that $\beta$ is essentially constant and reflects an intrinsic dispersion common to
all voters in a population, while the average of the city-dependent field $\mu$ depends strongly on the size of the city. Of course, all intermediate
scenarios are in principle possible too, but the data is not precise enough to hone in the precise relative contribution of the two effects. Here, we
want to argue that the dependence of $\mu$ on $N$ is likely to be dominant. Indeed, if the first scenario was correct, one should observe:
\be
m_N = \langle \tau \rangle_N \approx \langle \beta \rangle_N \Phi_{th}
\ee
The decrease of $m_N$ as a function of $N$ would therefore mean that $\langle \beta \rangle_N$ itself is a decreasing function of $N$ when the mean LTR is
positive. This is a priori reasonable: one expects more heterogeneity (and therefore a larger $\Sigma$, and a smaller $\beta$) in large cities than in small cities. 
However, the same model would imply a smaller dependence on $N$ for low turnout rates, and even an inverted dependence of $m_N$ on $N$ for elections with a very low turnover rate, 
such that $\langle \tau \rangle < 0$. 
This is not observed: quite on the contrary, the $m_N$ dependence is compatible with a mere vertical shift for similar elections, see Fig.~\ref{fmN-fr}.

On the other hand, a model where $\beta$ is constant, independent of $N$ and to a first approximation on the election, leads to:
\be
\langle \tau \rangle_N = m_N =  \beta \left[\Phi_{th} + \langle \mu \rangle_N\right],
\ee
which appears to be a good representation of reality. The dependence of $\langle \mu \rangle_N$ -- the average propensity to vote -- on $N$, could be the result of several intuitive mechanisms: 
for example, voters in small cities are less likely to be absent on election day (usually a sunday in France); the 
result of an election is sometimes more important in small cities than in large cities (for example, election of the mayor); the social pressure from
the rest of the community is stronger in small cities; all these effects suggest that the average turnout rate is stronger in small cities. In order 
to explain the opposite behaviour (as in Poland), or a non-monotonous dependence, as in Italy or Germany for parliament elections,  a systematic dependence of $\beta$ on $N$ might be 
relevant, although one should probably dwell into local idiosyncracies. 

Figure~\ref{fmN-fr} suggests that in France three families of elections clearly appear: a) ``important'' national elections (Presidential, Referendums, Parliament), for which $m_N$ shows a change of concavity around $N=1000$; b) less important national elections (European, {\it R\'egionales}) for which the average turnout is low, for which the change of concavity is absent; and c) {\it Municipales} for which the
variation of $m_N$ between small and large cities is the largest (as can be expected a priori). Note that the difference $\Delta m$ between the mean LTR for small and large cities is markedly different in the three cases: $\Delta m \approx 0.7$ in case a),  $\Delta m \approx 0.95$ in case b), and $\Delta m \approx 1.65$ in case c).

\begin{figure}
\begin{minipage}[c]{0.6\linewidth}
\includegraphics[width=9cm, height=6cm, clip=true]{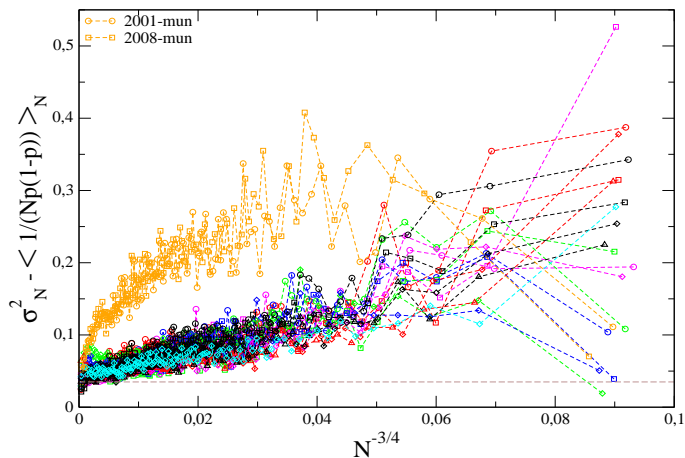}
\end{minipage}\hfill
\begin{minipage}[c]{0.35\linewidth}
\caption{\small $\sigma_N^2 - \langle \frac{1}{Np(1-p)} \rangle_N$ as a function of $N^{-3/4}$ for French elections. 
Each point comes from around 300 {\it communes} of size $\approx N$. Dashed line: $\beta^2\sigma^2_\phi \approx 0.035$ as extracted from the spatial correlations of $\tau$ (cf. Tab.~\ref{tcor-r}). 
The 1998 and 2004 {\it R\'egionales} elections are excluded here.}  
\label{fsigma2N-fr}
\end{minipage}
\end{figure}

As a first approximation, we thus take $\beta$ to be constant for all cities. The standard-deviation of $\tau$ 
over all cities of a given size then writes:
\be \label{sigmaN2}
\sigma_N^2 = \beta^2 \left[\langle \phi^2 \rangle + \langle \mu^2 \rangle_N - \langle \mu \rangle_N^2 \right] 
+ \frac{h}{N} \langle \frac{1}{p(1-p)} \rangle_N.
\ee
We show in Fig.~\ref{fsigma2N-fr} the quantity $\sigma_N^2$ minus the trivial binomial contribution, i.e. the last term of the RHS of the above equation, as a function 
of $N^{-3/4}$, for French elections. As predicted by the above model, we see that the $N \to \infty$ limit is clearly positive $\approx 0.035 \pm 0.05$, and to a good approximation 
independent of the election -- including the {\it Municipales}: although the $N$ dependence of $\sigma_N^2$ is found to be markedly different 
(as $N^{-1/4}$), this quantity still extrapolates to the
same asymptotic value. If one believes that our interpretation of $\phi$ as a persistent cultural field is correct, there is in fact
no reason to expect that $\sigma^2_\phi = \langle \phi^2 \rangle$ should change at all from election to election. The above result is therefore compatible with the fact that $\beta$ is to a first approximation 
election independent, as already suggested by Fig.~\ref{fmN-fr} above. The same results hold for all other countries, although the statistics is not as good as in the case of France: 
the asymptotic value of $\sigma_N^2$ for $N \to \infty$ is only weakly dependent on the election, and $\beta^2\sigma_\infty^2$ in the range $0.03 - 0.12$ for all countries. 
Furthermore, the $N$-dependence of $\sigma_N^2$ is found to be roughly compatible with $N^{-\omega}$ with $\omega \leq 1$ in all cases.

If $\beta$ is constant, the $N$-dependent contribution of $\sigma_N^2$ must come from the variance of the city-specific contribution $\mu$. A simple-minded model
for the statistics of $\mu$ predicts a variance that should decrease as $N^{-1}$. Indeed, a large city can be thought of as a patchwork of $n \propto N$
independent small neighbourhoods, each with a specific value of $\mu$. The effective value of $\mu$ for the whole city has a variance that is easily found to be reduced by a factor 
$n$, and therefore $\sigma_N^2 \propto N^{-1}$. A weaker dependence of $\sigma_N^2$ on $N$ signals the existence of strong inter-neighbourhood correlations (or strong heterogeneities in the
size of neighbourhoods), that lead to a reduction of the effective number of independent neighbourhood from $n \propto N$ to $n \propto N^\omega$ with $\omega < 1$. 
These inter-neighbourhood correlations are indeed expected, since some of the socio-economic and cultural factors affecting the decision of voters are clearly associated to the whole city. Interestingly, these correlations should be stronger for local elections, which is indeed confirmed by the fact that $\omega$ is markedly smaller for the {\it Municipales} elections in France. We therefore find the interpretation of the anomalous $N$ dependence of $\sigma_N^2$ as due to the city-specific contribution $\mu$ rather compelling. 

Let us now turn to the distribution of the rescaled variable $v$. Within the above model, and again assuming that $\beta$ is constant, one finds that:
\be
v = \frac{\tau - m_N}{\sigma_N} \propto \beta (\phi + \mu - \langle \mu \rangle_N) + \sqrt{\frac{h}{Np(1-p)}} \, \xi.
\ee
The last ``binomial'' term quickly becomes Gaussian as $N$ increases, and is at least four times smaller than 
the first two terms when $N > 1000$ (when $h=1$). Since the cultural field $\phi$ is, according to the model proposed in \cite{diffusive-field}, the result of averaging random influences over long time 
scales and large length scales, one expects, from the Central Limit Theorem, that $\phi$ is close to a Gaussian field as well. 
However, the statistics of $\mu$ has no reason to be Gaussian for small cities $N$, for which it
reflects local and instantaneous idiosyncracies, and for which no averaging argument can be invoked. The ``universality'' of $Q_N(v)$ across countries is therefore probably only apparent, 
since there is no 
reason to expect that the distribution of $\mu$ is independent of the country. In fact, $Q_N(v)$ in countries like Italy, Germany \& the Czech Republic do exhibit a stronger skewness than 
in other countries. Still, according to the above discussion, the contribution of different neighbourhoods to $\mu$ must average out as $N$ increases, and one expects the 
distribution of $\mu$ itself to become more and more Gaussian as $N$ increases. 

\begin{figure}
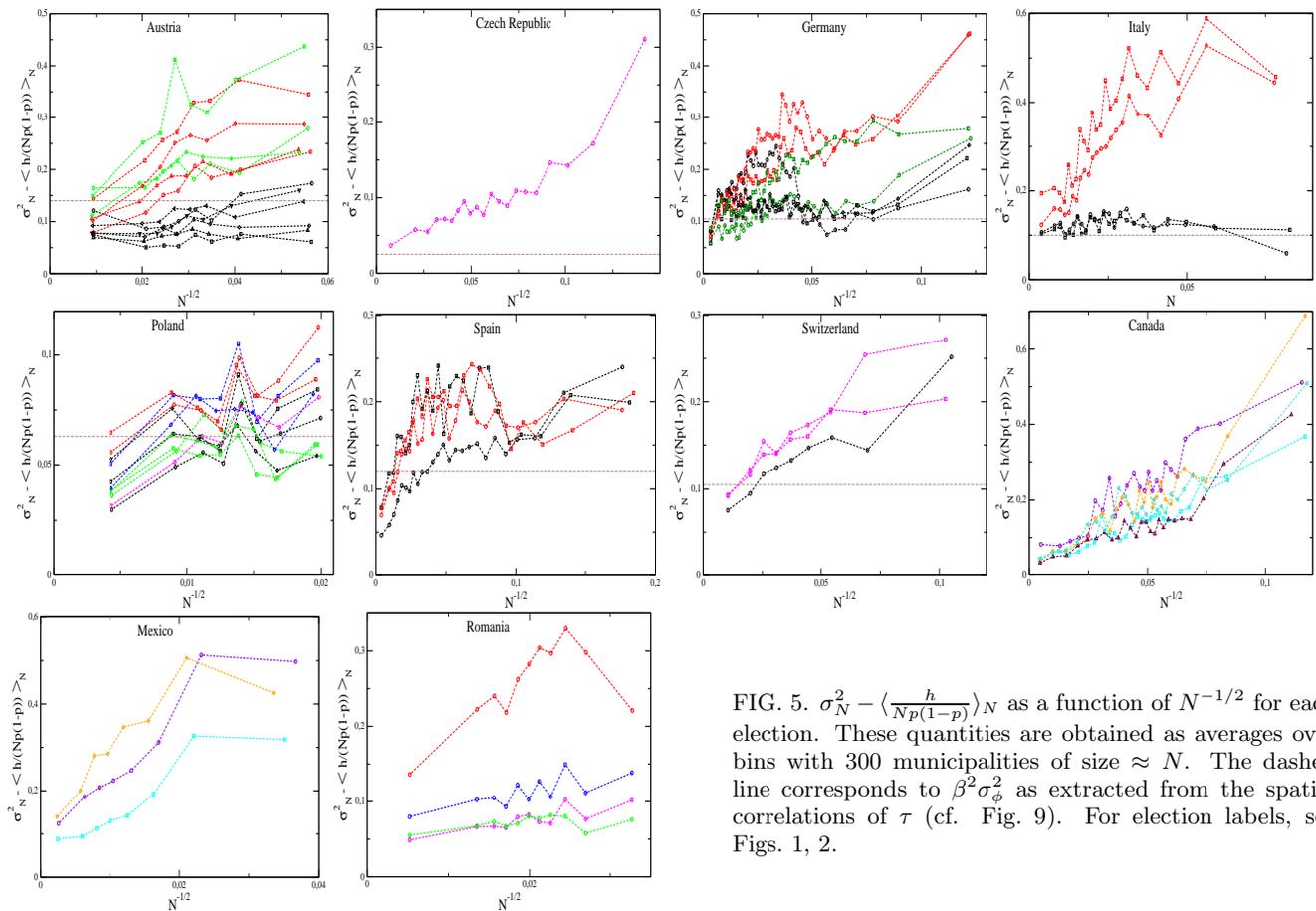

\includegraphics[width=4.25cm, height=4cm, clip=true]{sigma2N-at.eps}
\includegraphics[width=4.25cm, height=4cm, clip=true]{sigma2N-cz.eps}
\includegraphics[width=4.25cm, height=4cm, clip=true]{sigma2N-ge.eps}
\includegraphics[width=4.25cm, height=4cm, clip=true]{sigma2N-it.eps}\\
\includegraphics[width=4.25cm, height=4cm, clip=true]{sigma2N-pl.eps}
\includegraphics[width=4.25cm, height=4cm, clip=true]{sigma2N-sp.eps}
\includegraphics[width=4.25cm, height=4cm, clip=true]{sigma2N-ch.eps}
\includegraphics[width=4.25cm, height=4cm, clip=true]{sigma2N-ca.eps}
\begin{minipage}[c]{0.5\linewidth}
\includegraphics[width=4.25cm, height=4cm, clip=true]{sigma2N-mx.eps}
\includegraphics[width=4.25cm, height=4cm, clip=true]{sigma2N-ro.eps}
\end{minipage}\hfill
\begin{minipage}[c]{0.45\linewidth}
\caption{\small $\sigma_N^2 - \langle \frac{h}{Np(1-p)} \rangle_N$ as a function of $N^{-1/2}$ for each election. These quantities are obtained as averages over bins with 
300 municipalities of size $\approx N$. The dashed line corresponds to $\beta^2\sigma^2_\phi$ as extracted from the spatial correlations of $\tau$ (cf. Fig.~\ref{fcor-rescaled}). For election labels, see Figs.~\ref{fPu}, \ref{fmN}.}
\label{fsigma2N}
\end{minipage}
\end{figure}

To sum up: the random variable $v$ is the sum of three independent random variables, two of which can be 
considered as Gaussian, while the third has a distribution that depends on $N$ and becomes more Gaussian for large $N$, with a variance that decreases as $N^{-\omega}$. This allows one 
to rationalize the above empirical findings on the distributions $Q_N(v)$: these are more and more Gaussian as $N$ increases, and closer to one another for different countries, 
since the country specific contribution $\mu$ becomes smaller (as $N^{-\omega}$) and itself more Gaussian.

It is instructive to compare the relative contribution to the variance of the turnout rates of the cultural field $\phi$ on the one hand, and of the city-specific field on the other. The latter 
can be obtained by subtracting from the total variance of the LTR, $\sigma_\tau^2$, the contribution of the cultural field $\beta^2\sigma_\phi^2$ which is obtained as the extrapolation of $\sigma_N^2$ 
to $N \to \infty$ (see Figs \ref{fsigma2N-fr} \& \ref{fsigma2N}) and the average contribution of the binomial noise, $\langle h/Np(1-p) \rangle$. The herding factor $h$ can be estimated using the method introduced in
\cite{diffusive-field}, which compares different elections for which the binomial noises are by definition uncorrelated (see Eq. (10) of Ref. \cite{diffusive-field}). The ratio of $r=\sigma^2_\mu/\sigma_\phi^2$ can be seen as an objective  measure of the heterogeneity of behaviour in country, i.e. 
how strongly local idiosyncracies can depart from the global trend. Table~\ref{tcor-r} gives the ratio $r$ for all studied countries. Using this measure, we find that the most heterogeneous countries are 
Canada and the Czech Republic,\footnote{Although the ratios for Ca, Mx, Cz and Ge might be overestimated because the data did not allow us to estimate the herding ratio $h$ in these two cases.} and the most homogeneous ones are Austria, Switzerland and Romania. Not surprisingly, however, the largest value of $r$ is found for the French {\it Municipales}, i.e. local 
elections, for which idiosyncratic effects are indeed expected to be large. Note also that the herding ratio is anomalously high for Romania ($h=8.5$), and quite substantial for Poland ($h=4.7$).
Finally, it is interesting to notice that the quantity $\beta \sigma_\phi$ depends only weakly on the country (it varies by a factor $1.7$ between France and Italy). Since the total 
intention $\varphi$ is only defined up to an arbitrary scale, one can always set $\sigma_\phi=1$. Therefore, we find that the idiosyncratic dispersion $1/\beta$ (or the propensity not to conform to
the norm encoded by the cultural field) is strongest in France, Poland and the Czech Republic, and weakest in Italy and Austria.

\begin{table}[h!]
\begin{tabular}{|l|l|l|l|l|l|l|l|l|}
\hline
Country & $\sigma_\tau^2$ & $h$ & $\omega$ & $\beta^2\sigma_\phi^2$ ($N \to \infty$) & $\beta^2\sigma_\phi^2$ (Eq. \ref{rescale}) & $\langle h/(Np(1-p) \rangle$ & $\beta^2\sigma_\mu^2$ & $r=\sigma^2_\mu/\sigma^2_\phi$ \\
\hline
Fr	& $0.13$  & $0.8^\dagger$	& $3/4$  & $0.035$ & $0.035$ & $0.03$  &	$0.065$  & $1.85$ \\
Fr (mun)& $0.35$  & $1$                 & $1/4$  & $0.035$ & $0.035$ & $0.045$ &        $0.27$  &  $7.7$ \\
At	& $0.13$  & $2.9$	        & $1/2$  & $0.09$  & $0.14$  & $0.025$ &	$0.015$ &  $0.17$ \\
Pl	& $0.085$ & $4.7$	        & $1/2$  & $0.035$ & $0.065$ & $0.$    &	$0.05$  & $1.4$	\\
Ge	& $0.15$  & $0.^\star$	        & $1/4$  & $0.05$  & $0.105$ & $0.01$  &	$0.09$  & $1.8$ \\
Sp	& $0.195$ & $0.7^\dagger$	& $1/8$  & $0.06$  & $0.115$ & $0.035$ &	$0.10$  & $1.7$ \\
It	& $0.15$  & $2.2$	        & $1/4$  & $0.10$  & $0.10$  & $0.02$  &	$0.03$  & $0.3$ \\
CH	& $0.155$ & $0.6^\dagger$	& $1/2$  & $0.065$ & $0.105$ & $0.015$ &	$0.075$ & $0.85$\\
Cz	& $0.165$ & NA$^\diamondsuit$   & $1/2$  & $0.035$ & $0.035$ & $0.025$ &        $0.105$ & $3.$  \\
Ro      & $0.11$  & $8.5$               & $1/2$  & $0.07$  & NA      & $0.015$ &        $0.025$ & $0.36$ \\
Ca      & $0.2 $  & $1^\flat$           & $1/2$  & $0.03$  & NA      & $0.015$ &        $0.155$ & $5.1$ \\
Mx      & $0.27$  & $0.^\dagger$         & $1/2$  & $0.1 $  & NA     & $0.002$ &        $0.17$  & $1.7$ \\
\hline
\end{tabular}
\caption{\small Decomposition of the total LTR variance into a cultural field component $\beta^2\sigma_\phi^2$, and city-specific component $\beta^2\sigma_\mu^2$, and a binomial component, $\langle h/(Np(1-p) \rangle$, 
corrected by a herding coefficient $h \geq 1$. This last term is determined using the method proposed in \cite{diffusive-field}, which leads to a herding coefficient $h$
given in the second column. $\dagger$: when the direct fit gives a value of $h$ less than unity, we enforce $h=1$. $\star$: the case of Germany seems to be special, maybe due to a large fraction of
postal votes. $\diamondsuit$: the method to determine $h$ requires more than one election, and therefore cannot be applied to the Czech Republic. In this case, we also set $h=1$ by default. $\flat$:
Missing data prevents us from determining $h$ precisely, so we again set $h=1$ by default.
The value of the exponent $\omega$ is only indicative, since in some countries the power-law assumption is not warranted, see Fig.~\ref{fsigma2N}. We give two values for $\beta^2\sigma_\phi^2$: one as the 
asymptotic extrapolation of $\sigma_N^2 - \langle h/(Np(1-p) \rangle$ for $N \to \infty$ and the second from the rescaling coefficient $C^*$, see below and Fig.~\ref{fcor-rescaled}. 
Both these determinations are only precise to within roughly $\pm 20\%$.
}
\label{tcor-r}
\end{table}

\begin{figure}
\includegraphics[width=8.1cm, height=6cm, clip=true]{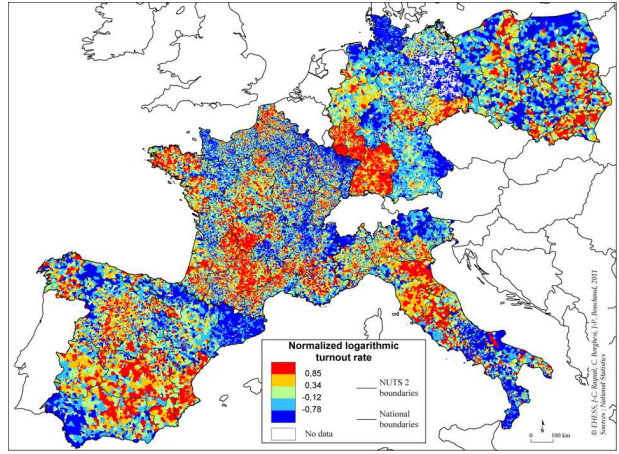}
\caption{\small Heat map of the normalized logarithmic turnout rate $\frac{\tau-m_N}{\sigma_N}$, for the 2004 European Parliament election in France, 
Germany, Italy, Poland and Spain. Germany had nomenclature reform of their municipalities which make more difficult to efficiently join spatial data to electoral data. 
Note the strongly heterogenous, but long-range correlated nature of the pattern. 
Note also some strong regionalities, for example in the German regions of Sarre or Bade-Wurtemberg, where the average turnout rate is strong and sharply falls across the
region boundaries. In these cases, the implicit assumption of a translation invariant statistical pattern that we make to compute $C(r)$ is probably not warranted, and 
it would in fact be better to treat these regions independently.
}
\label{fmap-Europe}
\end{figure}

\section{Spatial correlations of turnout rates}

\begin{figure}
\includegraphics[width=8.5cm, height=6cm, clip=true]{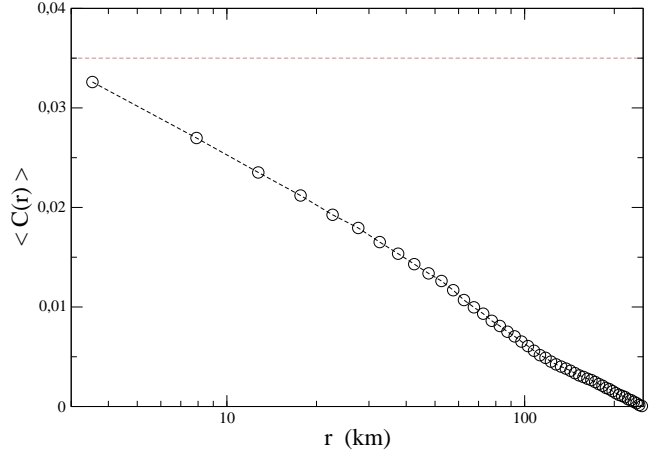}
\caption{\small Average of spatial correlations $C(r)$ for all French elections (absent the 2 {\it Municipales} elections). 
In dashed lines: $\beta^2 \sigma_\phi^2 \approx 0.035$, as extracted 
from the asymptotic ($N \to \infty$) dependence of $\sigma^2_N$.}
\label{fcov-France}
\end{figure}

\begin{figure}
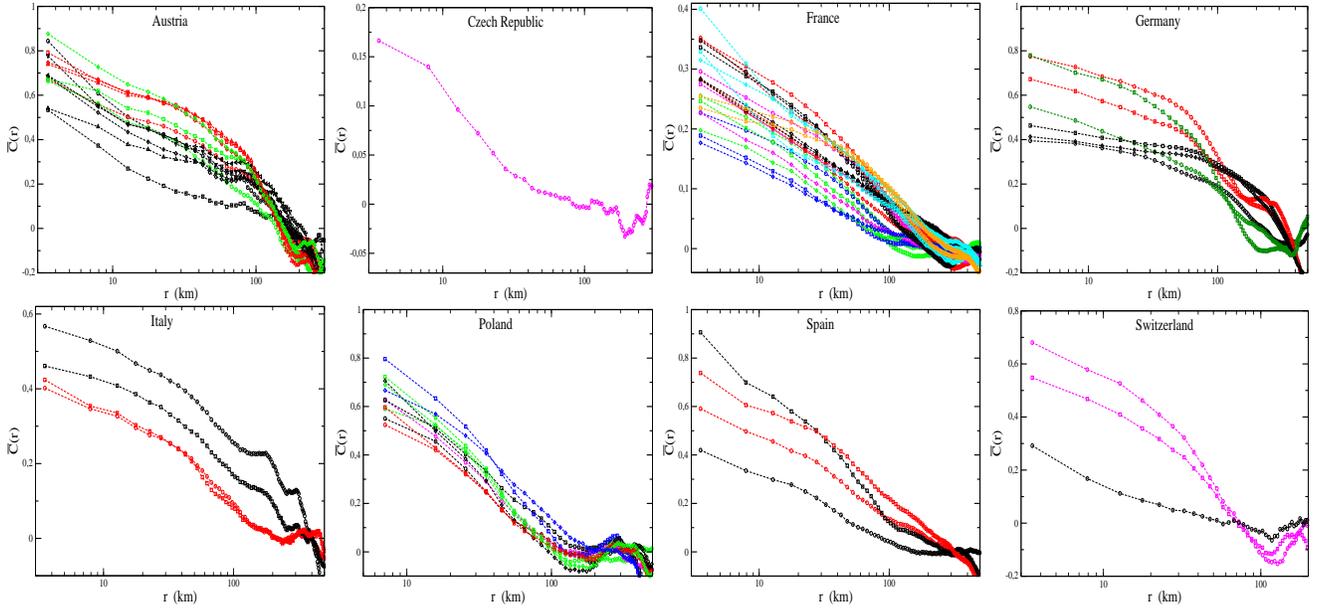

\includegraphics[width=4.25cm, height=4cm, clip=true]{cor-tau-at.eps}
\includegraphics[width=4.25cm, height=4cm, clip=true]{cor-tau-cz.eps}
\includegraphics[width=4.25cm, height=4cm, clip=true]{cor-tau-fr.eps}
\includegraphics[width=4.25cm, height=4cm, clip=true]{cor-tau-ge.eps}\\
\includegraphics[width=4.25cm, height=4cm, clip=true]{cor-tau-it.eps}
\includegraphics[width=4.25cm, height=4cm, clip=true]{cor-tau-pl.eps}
\includegraphics[width=4.25cm, height=4cm, clip=true]{cor-tau-sp.eps}
\includegraphics[width=4.25cm, height=4cm, clip=true]{cor-tau-ch.eps}
\caption{\small Normalized spatial correlations $\tilde C(r)$ of $\tau'=\tau - m_N$ for all countries for which the geographic position of cities is available. The correlation is normalized by the variance of $\tau'$, such that $\tilde C(r=0) \equiv 1$. For labels of elections, see Figs.~\ref{fPu}, \ref{fmN}.}
\label{fcor-r}
\end{figure}

Another striking empirical finding reported in \cite{diffusive-field,these} is the logarithmic dependence of the spatial correlation of the LTR as a function of distance. The spatial 
pattern of the local fluctuations of the LTR in European countrie are shown in Fig.~\ref{fmap-Europe}. One clearly sees the presence of long-ranged correlations. More precisely, for
the 13 French elections studied there, one finds that the spatial correlation of $\tau'(\vec R_\alpha)=\tau(\vec R_\alpha) - m_{N_\alpha}$ (where $\vec R_\alpha$ is the spatial location of the city and 
$m_{N}$ is the average of $\tau$ over cities of similar sizes) decreases as:
\be 
C(\vec r) = {\langle \tau'(\vec R + \vec r) \tau'(\vec R) \rangle}  \approx - C_0 \ln \frac{r}{L},
\ee
where $L$ is of the order of the size of the country. We show in Fig.~\ref{fcov-France} the average $C(r)$ for all French elections (except the two {\it Municipales} elections) and in Fig.~\ref{fcor-r}
the normalized correlation functions for all elections, separately for each country for which the geographic position of cities is available to us.

Using the above decomposition, and noting that by assumption the fluctuations of $\mu(\vec R)$ around the suitable size dependent average
$\langle \mu \rangle_N$ have {\it short-ranged correlations}, one concludes that the long-range, logarithmic correlations above must come from those of the cultural field $\phi$.
One indeed finds:
\be
C(\vec r \neq 0) \approx \langle \phi(\vec R + \vec r) \phi(\vec R) \rangle,
\ee
since the other two terms only contribute for $\vec r=0$. As a consistency check of this decomposition, one should find that $C(\vec r)$ should quickly decay from $C(r=0)$ to 
$C(r \to 0^+) \approx \beta^2 \sigma_\phi^2$ (e.g. $\approx 0.035 \pm 0.05$ for France). This is indeed seen to be well borne out, see Fig.~\ref{fcov-France}. The  agreement between two completely 
different determination of $\beta^2 \sigma_\phi^2$ (one using the extrapolation of $\sigma_N^2$ to infinite sizes, and the second using $C(r)$) 
holds very well for France, Italy and the Czech Republic, and only approximately for other countries (see Tab.~\ref{tcor-r} and Fig.~\ref{fsigma2N}).

Inspired by a well-known model in statistical physics where these logarithmic correlations appear, we postulated in \cite{diffusive-field} 
that the field $\phi$ evolves according to a diffusion equation, driven by a random noise, which is meant to describe the exchange of ideas and opinions between nearby cities and the random nature 
of the shocks that may affect the cultural substrate. As we argued in \cite{diffusive-field}, the fact that people move around and carry with them some components of the local cultural 
specificity leads to a local propagation of $\phi(\vec R_\alpha,t)$. Through human interactions, the cultural differences between nearby cities tend to narrow according to:
\be
\label{transfer}
\left.\frac{\partial \phi(\vec R_\alpha,t)}{\partial t}\right|_{\rm{infl.}} = \sum_\beta \Gamma_{\alpha\beta} [\phi(\vec R_\beta,t)-\phi(\vec R_\alpha,t)],
\ee
where $\Gamma_{\alpha\beta}(r_{\alpha\beta}) \geq 0$ is a symmetric influence matrix, that we assume to decrease over a distance corresponding to regular displacements of individuals, 
say $10$ km or so. For concreteness, we take:  $\Gamma_{\alpha\beta}(r)=\Gamma_0 e^{-r/\ell_c}$. As is well known, the continuum limit of the right hand side of Eq. (\ref{transfer}) reads $D \Delta \phi(\vec R,t)$, 
where $\Delta$ is the Laplacian and $D(\vec R_\alpha)=\frac12 \sum_{\beta} r_{\alpha\beta}^2 \Gamma_{\alpha\beta}$ is a measure of the speed at which the cultural field diffuses. 
Random cultural ``shocks'' add to the above equation a noise term $\eta(\vec R_\alpha,t)$. 

\begin{figure}
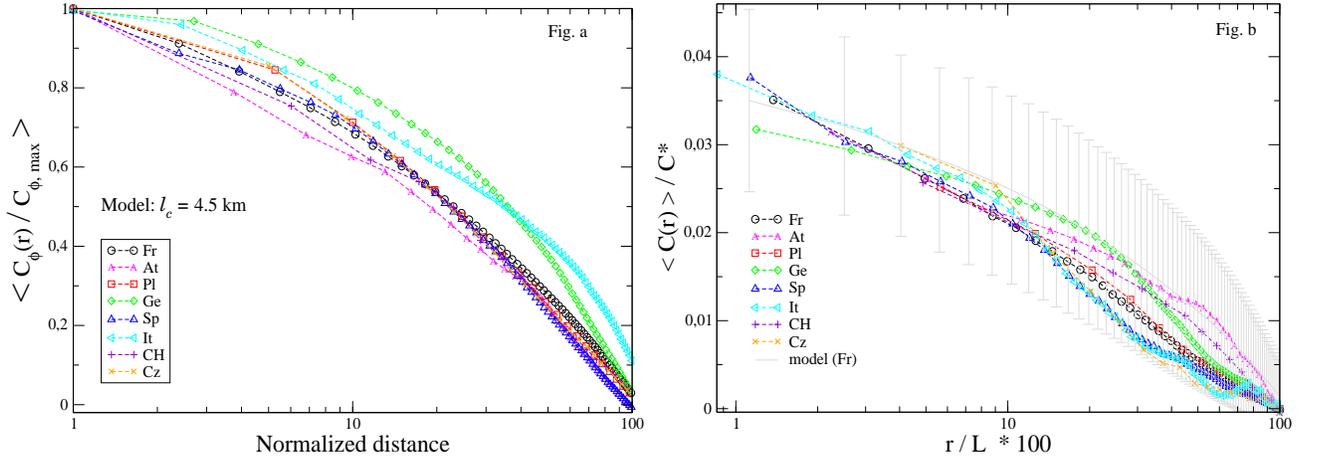

\includegraphics[width=8.5cm, height=6cm, clip=true]{cor-tau-norm-simul.eps}
\includegraphics[width=8.5cm, height=6cm, clip=true]{cov-moy-rescal.eps}
\caption{\small Average of spatial correlation, rescaled. 
Left: Average over numerical simulations of the model (with $\ell_c=4.5$~km) with the true positions of all cities for each country. 
Right: Average over real election data for each country. We also shown the average and standard deviation 
(coming from different realizations of the noise history $\eta$, and plotted as error bars) corresponding to the numerical model for French cities.}
\label{fcor-rescaled}
\end{figure}

If cities were located on the nodes of a regular lattice of linear size $L$, it would be easy to compute analytically the stationary correlation function of the field $\phi$. It is found to be given 
by a logarithm function of distance, provided $L \gg \ell_c$:
\be
C_\phi(r) \propto \ln \frac{L}{r}, \qquad \ell_c \ll r \ll L.
\ee
However, the spatial distribution of cities in real countries is quite strongly heterogeneous, which leads to significant deviation from a pure logarithmic decay. 
In order to compare quantitatively our model with empirical data, we have therefore simulated the model using Eq. (\ref{transfer}) with the {\it exact} locations of all cities for the
different countries under consideration. The results, averaged over many histories of the noise term, are shown in Fig.~\ref{fcor-rescaled}-left for $\ell_c=4.5$ km,
(but changing $\ell_c$ from $1.5$ km to $9$ km hardly changes the curves). Quite remarkably, we see that $C_\phi(r)$ exhibits a significant concavity, very similar to what is observed for 
the empirical correlations. In order to see that the model is indeed compatible with observations, we have plotted in Fig.~\ref{fcor-rescaled}-right the empirical data superimposed 
with the prediction of the model for the French case 
(for which the data is best). The empirical correlation $C(r)$ is rescaled by a country dependent value $C^*$ in order to achieve the best rescaling. This value of $C^*$ allows us to obtain a second 
determination of $\beta^2 \sigma_\phi^2$, through the relation:
\be
\beta^2 \sigma_\phi^2 = \left.\beta^2 \sigma_\phi^2 \right|_{Fr.} C^*.
\label{rescale}
\ee
Note however that the numerical model predicts a rather large dispersion around the average result, that comes from a strong dependence on the noise realisation $\eta(\vec R_\alpha,t)$. 
One should therefore expect that the empirical data (which corresponds to only a few histories) departs from the average theoretical curve, in a way perfectly compatible with 
Fig.~\ref{fcor-rescaled}-right. This also means that there is quite a bit of leeway in determination of $C^*$, which is only determined to within $\pm 20 \%$. Finally, note that the shape of 
$C(r)$ for Germany is significantly different, with a pronounced change of regime around $r \approx 70$ km. This is clearly related to the strong regional idiosyncracies that we discussed in 
Fig.~\ref{fmap-Europe}.

We conclude that our numerical model reproduces very satisfactorily the observations for {\it all studied countries} (with the possible exception of Germany, for the reason noted above). 
This lends strong support to the existence, conjectured in 
\cite{diffusive-field}, of an underlying diffusive cultural field responsible for both the long-range correlation (in space) and persistence (in time) of voting habits.

\section{Conclusion}

In this paper, we have shown that the empirical results for the statistics of turnout rates established in \cite{diffusive-field} for some French elections appear to hold much more generally.
We believe that the most striking result is the logarithmic dependence of the spatial correlations of these turnout rates. This result is quantitatively reproduced by a decision model that 
assumes that each voter makes his mind as a result of three influence terms: one totally idiosyncratic component, one city-specific term with short-ranged fluctuations in space, and one long-ranged
correlated field which propagates diffusively in space. The sum of these three contributions is what we call the ``intention''. A detailed analysis of our data sets has revealed several interesting 
(and sometimes unexpected) features: a) the city-specific term has a variance that depends on the size $N$ of the city as $N^{-\omega}$ with $\omega < 1$, suggesting strong inter-city correlations; b) different countries have different degrees of local heterogeneities, defined as the ratio of the variance of the city-dependent term over the variance of the cultural field; c) different countries seem to be characterized by a different propensity for individuals to conform to a cultural norm; d) there are clear signs of herding (i.e. strongly correlated decisions at the individual level) in 
some countries, but not in others; e) the statistics of the logarithmic turnout rates become more and more Gaussian as $N$ increases. 

Although we have confirmed the existence of a diffusive cultural field using election data from different countries, we feel that more work should be done to establish the general relevance of 
this idea to other decision making processes. It would be extremely interesting to find other data sets that would enable one to study the spatial correlations of decision making. An obvious 
candidate would be consumer habits -- for example the consumption pattern of some generic goods, or the success of some movie, etc.  

Finally, we believe that our detailed analysis of the statistics of turnout rates (or more generally of election results) reveals both stable patterns and subtle features, that could be used to 
test for possible data manipulation or frauds, or to define interesting ``democracy'' indexes. In that respect, the existence of strong herding effects in some countries is somewhat disturbing.  

\section{Materials and Methods}
The Appendix gives more information about the set of (public) electoral data studied in this paper. Most of them can be directly downloaded from official websites (see References).

Average values and standard-deviations do not take into account extreme values in order to remove some electoral errors, etc. Electoral values greater than \textit{5 sigma} are not taken into account~\footnote{For instance let 100 municipalities of size $\approx N$ (as in Fig.~\ref{fmN}), each one has a LTR $\tau_i$ ($i=1,2,...,100$). First, $\langle\tau\rangle$ and $\sigma$ are the average value and the standard-deviation of $\tau$ over these 100 municipalities. Next, the final average value $m_N$ and the final standard-deviation, $\sigma_N$, over this sample of 100 municipalities are uniquely evaluated for municipalities, $i$, such that $|\tau_i - \langle\tau\rangle| < 5\;\sigma$.}.

\section*{Appendix: Details on the data sources}
Table~\ref{tdata} shows the nature of the 77 national elections from 11 countries, studied at the municipality scale. Countries are: France (Fr)\footnote{1994 and 2004 {\it R\'egionales} elections occurred at the same time as strictly local elections ({\it cantonales}, i.e. at a kind of county level) in half of municipalities.}, Austria (At)\footnote{Postal votes ({\it Wahlkarten}) are not taking account in this paper.}, Poland (Pl), Germany (Ge)\footnote{{\it L\"and} Parliament elections at time less or equal to 2004 (or 2010) in each {\it Land} are written here as `2004-Ld' (or `2010-Ld').}, Spain (Sp), Italy (It), Swiss (CH)\footnote{The referendums or \textit{votations} ($\mathrm{R_{(a)}}$ and $\mathrm{R_{(b)}}$) respectively occurred on March 11th and July 17th 2007.}, Czech Republic (Cz), Canada (Ca), Romania (Ro)\footnote{The referendum studied here (about the \textit{Parlament unicameral} and the reduction of the maximum of deputies) occurred at the same time than the first round of the Presidential election. Some Romanian electors, not registered in the \textit{lista electorala permanenta}, are able to vote. For this country, we pursue to write $N$ the Number of Register Voters, $N_+$ the registered electors who take part to the election.} and Mexico (Mx). Note that all the studied elections occurred in a same time over all the country (apart from 2 \textit{L\"ander} elections in Germany) and are free of compulsory voting. Lastly, in our database for Germany, postal votes ({\it Briehwahlen}) are taken into account in some {\it L\"ander}, not in others, which artificially increases turnout heterogeneity between German regions.

Moreover Election turnout statistics have been located, identified and geocoded, based on a set of points, which were obtained by calculating the gravity center of each municipality or the position of the town-hall, and then adding the X and Y coordinates for each of these features. In addition to these coordinates, the objects are described with several attributes: logarithmic turnout rate, $\tau$, normalized logarithmic turnout rate, $v$, etc. This concerns 8 countries amongst the 11 previous ones~\footnote{The Mexican spatial repartition of municipalities is so widely heterogeneous than the spatial study made for other countries is no longer efficient here.}: Austria~\cite{XY-at}, Czech Republic~\cite{XY-cz}, France~\cite{XY-fr}, Germany~\cite{XY-ge}, Italy~\cite{XY-it}, Poland~\cite{XY-pl}, Spain~\cite{XY-sp} and Switzerland~\cite{XY-ch}. This study is limited to mainland municipalities (and each considered country have more than two thousands municipalities). {\it Lambert 2 \'etendu} is used for France, while {\it WGS~84} coordinate system is used for other countries.

\begin{table}[h!]
\begin{tabular}{| l l l l l |}
\hline
Ctry & $\#$el & mun & spa & elections\\
\hline
\multirow{2}*{Fr} & \multirow{2}*{22} & \multirow{2}*{36000} & \multirow{2}*{Y} & 1992-R, 1993-D, 1994-E, 1995-P1, 1995-P2, 1997-D, 1998-rg, 1999-E, 2000-R, 2001-mun, 2002-P1, \\
 & & & & 2002-P2, 2002-D, 2004-rg, 2004-E, 2005-R, 2007-P1, 2007-P2, 2007-D, 2008-mun, 2009-E, 2010-rg \\
At & 13 & 2400 & Y & 1994-D, 1995-D, 1996-E, 1998-P, 1999-E, 1999-D, 2002-D, 2004-P, 2004-E, 2006-D, 2008-D, 2009-E, 2010-P \\
Pl & 11 & 2500 & Y & 2000-P1, 2001-D, 2003-R, 2004-E, 2005-D, 2005-P1, 2005-P2, 2007-D, 2009-E, 2010-P1, 2010-P2 \\
Ge & 7 & 12000 & Y & 2002-D, 2004-Ld, 2005-D, 2009-E, 2009-D, 2010-Ld \\
Sp & 4 & 8000 & Y & 2004-D, 2004-E, 2008-D, 2009-E \\
It & 4 & 7200 & Y & 2004-E, 2006-D, 2008-D, 2009-E \\
CH & 3 & 2700 & Y & 2007-R$_{(1)}$, 2007-R$_{(2)}$, 2007-D \\
Cz & 1 & 6200 & Y & 2003-R \\
Ca & 5 & 7700 & N & 1997-D, 2000-D, 2004-D, 2006-D, 2008-D \\
Ro & 4 & 3200 & N & 2009-E, 2009-R, 2009-P1, 2009-P2 \\
Mx & 3 & 2400 & N & 2003-D, 2006-D, 2009-D \\
\hline
\end{tabular}
\caption{Nature of elections studied in this paper. For each country (Ctry), the number of elections ($\#$el) and the number of municipalities(mun) in the mainland are written. "Y" (or reversely "N") mentions that municipalities are spatially (spa) localized. For each country, an election is identified by its year date and its nature. D: Chamber of Deputies election; E: European parliament election; P: presidential election (according to the constitution of the country, in only one round); P1 and P2: first and second round of a Presidential election; R: Referendum; Ld: German \textit{L\"ander} elections; rg: French \textit{R\'egionales} elections; mun: French {\it municipales}. For each country elections are given in a chronological order (but the 2009 Romanian Presidential (P) and Referendum (R) elections occurred the same day). Even if an election needs two rounds, only the first one is considered (e.g. the French Chamber of Deputies (D), \textit{R\'egionales} (rg) and {\it municipales} (mun) elections) unless the contrary is indicated (e.g. P1 and P2).}
\label{tdata}
\end{table}

\section*{acknowledgments}
C. B. would like to thank Brigitte Hazart, from the French \textit{Minist\`ere de l'Int\'erieur, bureau des \'elections et des \'etudes politiques}; Nicola A. D'Amelio, from the Italian \textit{Ministero dell'interno, Direzione centrale dei servizi elettorali}; Radka Sm\'idov\'a, from the \textit{Czech Statistical Office, Provision of electronic outputs}; Claude Maier and Madeleine Schneider, from the Swiss \textit{Office f\'ed\'eral de la statistique, Section Politique, Culture et M\'edias}; Alejandro Vergara Torres, Antonia Ch\'avez, from the Mexican \textit{Instituto Federal Electoral}; Matthias Klumpe from the German \textit{Amt f\"ur Statistik Berlin-Brandenburg}; anonymous correspondents of the \textit{\'Elections Canada, Centre de renseignements} and of the Spanish \textit{Ministerio del Interior, Subdirecci\'on General de Pol\'itica Interior y Procesos Electorales}, for their explanations and also for the great work they did to gather and make available the electoral data that they sent us.

\end{document}